\documentclass[manuscript]{aastex}
\usepackage{pslatex,times,emulateapj5}

\newcommand{\civ}{C\,{\sc iv}}
\newcommand{\ebv}{$E(B-V)$}
\newcommand{\etal}{{\it et\,al.}}
\newcommand{\ha}{H$\alpha$}
\newcommand{\hi}{H\,{\sc i}}

\newcommand{\ho}{$H_0$}
\newcommand{\kms}{km\,s$^{-1}$}
\newcommand{\kmsm}{km\,s$^{-1}$\,Mpc$^{-1}$}

\newcommand{\mgii}{Mg\,{\sc ii}}

\newcommand{\aliii}{Al\,{\sc iii}}

\newcommand{\cii}{C\,{\sc ii}}

\newcommand{\ciii}{C\,{\sc iii}]}

\newcommand{\feii}{Fe\,{\sc ii}}

\newcommand{\heii}{He\,{\sc ii}}

\newcommand{\lya}{Ly$\alpha$}

\newcommand{\Nv}{N\,{\sc v}}

\newcommand{\NIii}{Ni\,{\sc ii}}
\newcommand{\oi}{O\,{\sc i}}

\newcommand{\Oiii}{O\,{\sc iii}}

\newcommand{\Siiii}{Si\,{\sc iii}]}
\newcommand{\SIiv}{Si\,{\sc iv}}

\newcommand{\al}{$\alpha_{\lambda}$}
\newcommand{\pg}{PG 1407+265}
\newcommand{\sdssj}{SDSS J1136+0242}

\shortauthors{Hall \etal}
\shorttitle{LY-$\alpha$-ONLY AGN}

\begin{document}


\journalinfo{submitted to AJ 12/22/2003; revised version submitted 02/24/2004; accepted 02/25/2004 for the June 2004 issue}
\submitted{}

\title{A Lyman-$\alpha$-only AGN from the Sloan Digital Sky Survey}

\author{
Patrick B. Hall,\altaffilmark{1,2,3,4,5}
Erik A. Hoversten,\altaffilmark{6}
Christy A. Tremonti,\altaffilmark{7}
Daniel E. Vanden Berk,\altaffilmark{8}
Donald P. Schneider,\altaffilmark{9}
Michael A. Strauss,\altaffilmark{1}
Gillian R. Knapp,\altaffilmark{1}
Donald G. York,\altaffilmark{10}
Damien Hutsem\'ekers,\altaffilmark{5,11}
P. R. Newman,\altaffilmark{12}
J. Brinkmann,\altaffilmark{12}
Brenda Frye,\altaffilmark{1}
Masataka Fukugita,\altaffilmark{13}
Karl Glazebrook,\altaffilmark{6}
Michael Harvanek,\altaffilmark{12}
Timothy M. Heckman,\altaffilmark{6}
\v{Z}eljko Ivezi\'{c},\altaffilmark{1}
S. Kleinman,\altaffilmark{12}
Jurek Krzesinski,\altaffilmark{12,14}
Daniel C. Long,\altaffilmark{12}
Eric Neilsen,\altaffilmark{15}
Martin Niederste-Ostholt,\altaffilmark{1}
Atsuko Nitta,\altaffilmark{12}
David J. Schlegel,\altaffilmark{1}
S. Snedden\altaffilmark{12}
}
\altaffiltext{1}{Princeton University Observatory, Princeton, NJ 08544-1001}
\altaffiltext{2}{Departamento de Astronom\'{\i}a y Astrof\'{\i}sica, 
Facultad de F\'{\i}sica, Pontificia Universidad Cat\'{o}lica de Chile, 
Casilla 306, Santiago 22, Chile; E-mail: phall@astro.puc.cl}
\altaffiltext{3}{Guest Observer at the Cerro Tololo Inter-American Observatory,
a division of the National Optical Astronomy Observatories, which is operated 
by AURA, Inc. under cooperative agreement with the National Science Foundation.}
\altaffiltext{4}{Guest Observer at the Canada-France-Hawaii Telescope,
a joint facility of the National Research Council of Canada, the Centre
National de la Recherche Scientifique of France and the University of Hawaii.}
\altaffiltext{5}{Guest Observer at the La Silla Observatory ESO 3.6m Telescope,
for program 071.B-0460.}
\altaffiltext{6}{Department of Physics and Astronomy, The Johns Hopkins
University, 3400 North Charles Street, Baltimore, MD 21218-2686}
\altaffiltext{7}{Steward Observatory, The University of Arizona, 
933 North Cherry Avenue, Tucson AZ 85721}
\altaffiltext{8}{Department of Physics and Astronomy, University of Pittsburgh,
3941 O'Hara Street, Pittsburgh, PA 15260}
\altaffiltext{9}{Department of Astronomy and Astrophysics, 
The Pennsylvania State University, University Park, PA 16802}
\altaffiltext{10}{Department of Astronomy and Astrophysics and Enrico Fermi
Institute, The University of Chicago, 5640 S. Ellis Ave., Chicago, IL 60637}
\altaffiltext{11}{Research Associate FNRS, University of Li\`ege,
     All\'ee du 6 ao\^ut 17, Bat. 5c, 4000 Li\`ege, Belgium}
\altaffiltext{12}{Apache Point Observatory, P.O. Box 59, Sunspot, NM 88349-0059}
\altaffiltext{13}{University of Tokyo, Institute for Cosmic Ray Research, 
Kashiwa, 2778582, Japan}
\altaffiltext{14}{Mt. Suhora Observatory, Cracow Pedagogical University, 
ul. Podchorazych 2, 30-084 Cracow, Poland}
\altaffiltext{15}{Fermi National Accelerator Laboratory, 
P.O. Box 500, Batavia, IL 60510}

\begin{abstract}
The Sloan Digital Sky Survey has discovered a $z=2.4917$ radio-loud active
galactic nucleus (AGN) with a luminous, variable, low-polarization UV continuum,
\hi\ two-photon emission, and a moderately broad \lya\ line
(FWHM$\simeq$1430\,\kms) but without obvious metal-line emission.  
SDSS J113658.36+024220.1 does have associated metal-line absorption
in three distinct, narrow systems spanning a velocity range of 2710 \kms.
Despite certain spectral similarities, \sdssj\ is not a Lyman-break galaxy.
Instead, the \lya\ and two-photon emission can be attributed to an extended,
low-metallicity narrow-line region.
The unpolarized continuum argues that we see \sdssj\ 
very close to the axis of any ionization cone present.  We can conceive of
two plausible explanations for why we see a strong UV continuum
but no broad-line emission in this `face-on radio galaxy' model for \sdssj: the
continuum could be relativistically beamed synchrotron emission which swamps the
broad-line emission; or, more likely, \sdssj\ could be similar to \pg, a quasar
in which for some unknown reason the high-ionization
emission lines are very broad, very weak, and highly blueshifted.
\end{abstract}

\section{Introduction}  \label{INTRO}

One of the goals of the Sloan Digital Sky Survey \markcite{yor00}(SDSS; {York} {et~al.} 2000)
is to obtain spectra for $\sim$10$^5$ quasars,
in addition to the $\sim$10$^6$ galaxies which comprise the bulk of the
spectroscopic targets \markcite{str02,sdss89}({Strauss} {et~al.} 2002; {Blanton} {et~al.} 2003).  From astrometrically calibrated
drift-scanned imaging data \markcite{gun98,sdss153}({Gunn} {et~al.} 1998; {Pier} {et~al.} 2003) on the SDSS $ugriz$
AB asinh magnitude system \markcite{fuk96,sdss26,sdss82,sdss85,sdss105}({Fukugita} {et~al.} 1996; {Lupton}, {Gunn}, \& {Szalay} 1999; {Hogg} {et~al.} 2001; {Stoughton} {et~al.} 2002; {Smith} {et~al.} 2002),
quasar candidates are selected primarily using color criteria designed to
target objects whose broad-band colors differ from those of normal stars
and galaxies \markcite{sdssqtarget}({Richards} {et~al.} 2002a).  
Some of these quasars are bound to possess unusual properties.

Here we describe one such unusual quasar,
SDSS J113658.36+024220.1 (hereafter SDSS J1136+0242),
which exhibits strong \lya\ emission but no detected metal-line emission.
We discuss its unusual properties (\S \ref{SPEC}),
consider possible explanations for them (\S \ref{WASSUP}),
and conclude with suggestions for further investigation (\S \ref{CON}).
Where needed, we assume \ho=70 \kmsm, $\Omega_M$=0.7 and $\Omega_{\Lambda}$=0.3.

\section{Observations and Properties of SDSS J1136+0242} \label{SPEC}

\subsection{Photometry and Variability} \label{PHOTVAR}

A catalog of observations and general properties of \sdssj\ is given in Table
\ref{t_info}.  Besides the SDSS imaging and spectroscopy,
near-IR photometry was obtained on UT 2003 January 31 using the
Infrared Side Port Imager (ISPI; \markcite{ispi}{Probst} {et~al.} 2003) on the CTIO 4-meter telescope,
with total exposure times of 800 seconds in $J$ and 300 seconds in $K_s$.

This object was imaged twice by the SDSS, and was unresolved in both epochs.
A $\chi^2$ analysis confirms variability between the two epochs of SDSS $ugriz$
imaging data (with $\Delta t$=67 rest-frame days) at the 99.8\%
confidence level.  Similarly, $gri$ magnitudes synthesized from 
the spectroscopy differ at the 85\% confidence level 
from the earlier SDSS imaging magnitudes ($\Delta t$=42 rest-frame days),
but not from the later ones ($\Delta t$=25 rest-frame days).
All imaging magnitudes come from data reduced with the same version of the SDSS
photometric pipeline, and the spectroscopic comparison is empirically
calibrated using observations of non-variable stars \markcite{vdb04}({Vanden Berk et al.} 2004).

\subsection{Polarimetry} \label{POL}

Photopolarimetry of \sdssj\ was obtained on UT 2003 April 5
using the EFOSC instrument on the ESO 3.6m telescope at La Silla Observatory.
\sdssj\ is unpolarized, with observed $i$-band (rest-frame $\sim$2300~\AA)
$P=0.7\pm0.4$\% at $\theta$=155$\pm$19\arcdeg\ E of N.
The observed lack of polarization means that the observed continuum 
does not come from an elongated region surrounding the central source.

\subsection{Radio Emission} \label{RADIO}

\sdssj\ is a 1.44 mJy radio source
in the FIRST survey \markcite{bwh95}({Becker}, {White}, \& {Helfand} 1995).  
Despite being near the FIRST flux limit, it appears marginally resolved in one
dimension, with a deconvolved major axis extent of 6\farcs84 at position angle
74.4$\pm$3.9\arcdeg\ E of N.  The FIRST peak flux density is only 1.01 mJy,
which is also suggestive of an extended source, though not conclusively so
\markcite{sdss1st}({Ivezi{\' c}} {et~al.} 2002).  According to
NED,\footnote{The NASA/IPAC Extragalactic Database (NED) is operated by the 
Jet Propulsion Laboratory, California Institute of Technology, under contract
to NASA.} \sdssj\ is not detected in any radio catalog besides FIRST; thus,
no interesting constraint can be placed on its radio spectral index.

\subsection{Spectroscopy and Redshift Determination} \label{SPECZ}

\begin{figure}
\epsscale{0.855}
\plotone{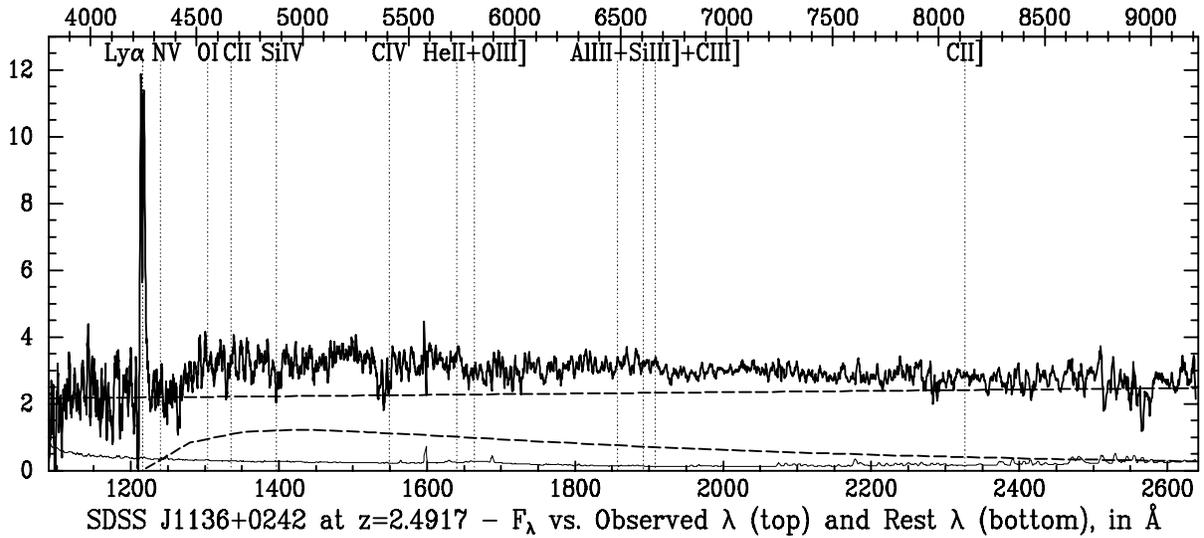}
\caption[]{
\singlespace 
Full SDSS spectrum (resolution $R\simeq2000$) of SDSS J1136+0242, smoothed by a
7-pixel boxcar filter, and accompanying 1$\sigma$ uncertainties.  The ordinate
is $F_{\lambda}$ in units of 10$^{-17}$ ergs cm$^{-2}$ s$^{-1}$ \AA$^{-1}$.
Wavelengths of emission lines which are often strong in AGNs
are marked, but only \lya\ is unambiguously detected.  
The dashed lines show the decomposition of the continuum into a power
law with $F_{\lambda} \propto \lambda^{0.15}$ plus \hi\ two-photon continuum.
}\label{f_full}
\end{figure}

\sdssj\ was targeted for spectroscopy in the SDSS as a {\sc serendipity} target
(a category including FIRST sources fainter than $i=19.1$; \markcite{sdss85}{Stoughton} {et~al.} 2002)
and as a high redshift quasar candidate.\footnote{Note that using 
the final version of the SDSS targeting software, \sdssj\ no longer qualifies
as a high-redshift quasar candidate.  \sdssj\ has colors red enough to place it
very close to the stellar locus, and the SDSS quasar targeting algorithms are
necessarily
incomplete to such objects to avoid being swamped by false positives.  Objects
with lines as weak as \sdssj\ but with bluer (or much redder) continua will
be targeted, but the SDSS is not complete to objects very similar to \sdssj.
}
The SDSS spectrum (Fig. \ref{f_full}) has a median signal-to-noise ratio (SNR)
per pixel ranging from 4 in the $g$ band to 7 in the $i$ band.
It shows a single, moderately broad \lya\ emission line (FWHM$\simeq$1430 \kms) 
with accompanying \lya\ and \civ\ absorption at three distinct redshifts
within a velocity range of 2710 \kms.
The curvature in the continuum between \lya\ and $\sim$1280~\AA\ is almost
certainly due to two-photon emission from the $2^2S$ state of \hi\ (see 
\S\ref{2Q}).  The spectrum is well fit by the sum of two-photon
emission and a power-law with $F_{\lambda}\propto\lambda^{0.15}$ (dashed lines;
reduced chi-squared is $\chi_{\nu}^2=1.106$,
neglecting systematic errors in the spectrophotometry).
This power-law spectral index is much redder than 
the \al$=-0.4$ found by \markcite{sea02}{Steidel} {et~al.} (2002)
for a sample of narrow-line, radio-quiet AGNs at $z\simeq3$
or the \al$=-1.56$ found for the typical SDSS quasar 
\markcite{sdss73}({Vanden Berk} {et~al.} 2001).
In fact, less than 0.5\% of SDSS quasars in the study of 
\markcite{redqso}{Richards} {et~al.} (2003) have such red continua.

\begin{figure}
\epsscale{0.855}
\plotone{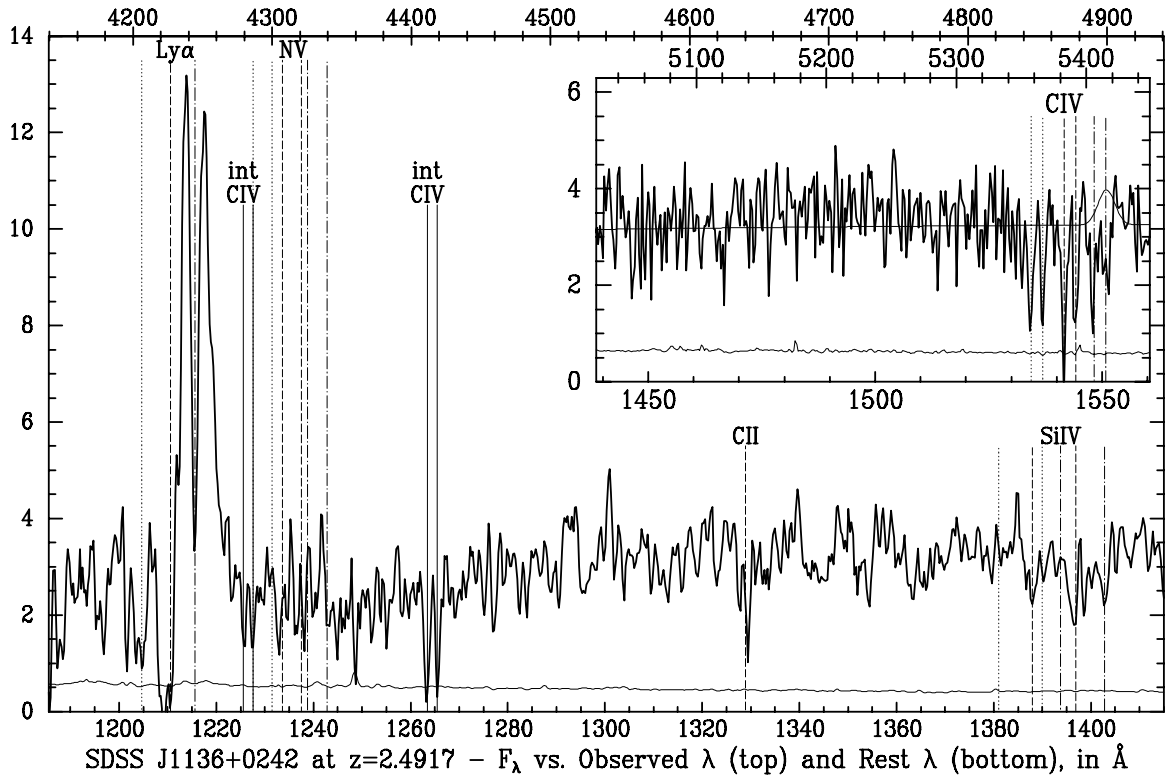}
\caption[]{
\singlespace 
SDSS spectrum and 1$\sigma$ uncertainties 
in regions of strong associated absorption.  The ordinates are
$F_{\lambda}$ in units of 10$^{-17}$ ergs cm$^{-2}$ s$^{-1}$ \AA$^{-1}$.
The three associated systems have \civ\ redshifts of
$z=2.4604$ (dotted), $z=2.4768$ (dashed), and $z=2.4917$ (dot-dashed).
The vertical solid lines show intervening \civ\ systems.
The inset shows the unsmoothed spectrum around \civ, along with
the best-fit continuum and variable-$\lambda_{cen}$ \civ\ emission line.
The main panel shows the spectrum from \lya\ to \SIiv,
smoothed by a 3-pixel boxcar.
\SIiv\ absorption (see Table 1) is most reliably detected at $z=2.4768$
(dashed), and \cii\ as well,
although the latter is at a slightly higher redshift of $z=2.4784$.
\Nv\ is not detected at any of the three associated absorption redshifts,
nor does it appear in emission.
}\label{f_zoom}
\end{figure}

To determine the systemic redshift of \sdssj, we simultaneously fitted for
the \lya\ emission redshift, the \lya\ absorption redshifts,
and the \civ\ absorption redshifts using IRAF.\footnote{The Image 
Reduction and Analysis Facility (IRAF) 
is distributed by the National Optical Astronomy Observatories, which is
operated by AURA, Inc., under contract to the National Science Foundation.}
We assumed a single Gaussian profile for each component.
Each line's width, height and center was allowed to vary independently.
The \lya\ absorption redshifts very closely matched the \civ\ absorption
redshifts (Fig. \ref{f_zoom}), demonstrating that the \lya\ emission does 
not bias the determination of the \lya\ absorption systems' parameters.
The best-fit \lya\ emission redshift was identical
to that of the highest redshift absorption system to within the errors.
We therefore adopted the redshift of this system,
$z=2.4917\pm0.0009$, as the systemic redshift.
(The other associated absorption systems are at $z=2.4603\pm0.0004$ and 
$z=2.4768\pm0.0002$.)

To place limits on emission line fluxes, we simultaneously fitted all
broad emission and narrow absorption features
listed in Table \ref{t_spec} to the rest-frame spectrum.  
The spectrum was first normalized by a single third-order spline fit to
rest-frame spectral regions at 1090--2460~\AA\ free of strong emission
lines seen in normal quasars (inferred from Table 2 of \markcite{sdss73}{Vanden Berk} {et~al.} 2001).
Narrow absorption lines were
allowed to vary in central wavelength, depth, and width, except as noted.
Because of the obvious weakness of any emission lines besides \lya, all broad
emission lines were constrained to have the same width as \lya.
For \Nv\ and \civ\ doublet emission, we fit a single gaussian whose FWHM was
the quadrature sum of the \lya\ FWHM and the separation between the two doublet
lines. 
Two fits to the emission lines were made:
one where the central wavelength $\lambda_{cen}$ was allowed to vary 
because AGN emission lines often exhibit velocity shifts \markcite{emshift}({Richards} {et~al.} 2002b),
and one where $\lambda_{cen}$ was held fixed at the $\lambda_{obs}$ values 
from Table 2 of \markcite{sdss73}{Vanden Berk} {et~al.} (2001).
We refer to those fits as the {\em variable-$\lambda_{cen}$} 
and {\em fixed-$\lambda_{cen}$} fits, respectively.
Only the variable-$\lambda_{cen}$ fit results are listed in Table \ref{t_spec},
but line ratios for both fits are listed in Table \ref{t_ratio}
(see Appendix \ref{BLAGN}).

As seen in Table \ref{t_spec}, a few emission lines are nominally
detected with rest-frame equivalent width (REW) values which are non-zero at 
$\gtrsim 3\sigma$ significance.  These $\sigma$ values do not include
the uncertainties in the continuum fit, and so may be underestimates;
however, we have also ignored the fact that the errors in the FWHM and the
peak value are correlated, which will make the $\sigma$ values overestimates.
The differences between the fluxes from the variable-$\lambda_{cen}$ and 
fixed-$\lambda_{cen}$ fits are illustrative of the systematic uncertainties.

Because there is \civ\ absorption at the systemic redshift, the fit to the 
\civ\ emission line depends on the fit to the absorption.  
An unconstrained fit to the \civ\ spectral region yields considerable broad
emission which is masked by relatively broad absorption (FWHM of 574 \kms).
While we cannot completely rule out such a fit, we adopt a more plausible fit
where the \civ\ absorption lines are constrained to have the same FHWM as the
\lya\ absorption at the same redshift (431 \kms).  Even that width is probably
an overestimate for the \civ\ absorption, as the other two associated systems
have \civ\ widths only 20$-$40\% as broad as \lya.  Nonetheless, this fit is
conservative in the sense that narrower widths would further reduce the 
allowable \civ\ emission strength, moving this object even further from the
norm for quasars.

\subsection{Hydrogen Two-Photon Emission and the Overall Spectrum} \label{2Q}

The curvature in the continuum of \sdssj\ 
between \lya\ and $\sim$1280~\AA\ matches the
characteristic spectral shape of forbidden two-photon emission from the
$2^2S$ state of hydrogen \markcite{all84,ost89}({Aller} 1984; {Osterbrock} 1989).
The spectrum is well fit by the sum of two-photon emission and a power-law
with $F_{\lambda} \propto \lambda^{0.15}$, and
the SDSS photometry from the same epoch as the spectroscopy is also consistent
with such a spectrum.\footnote{The photometry from the earlier photometric 
epoch, when the object was fainter, is better fit with a redder power-law index
($\alpha_\lambda=0.30$).}
The $J-K_s$ color corresponds to a power-law spectrum of index
$\alpha_\lambda=-0.14\pm0.30$, consistent within the uncertainties with an
extrapolation of the value of $\alpha_\lambda=0.15$ that fits the SDSS spectrum.
The $J$ and $K_s$ magnitudes are fainter than an extrapolation of such a
power-law; however, given the likelihood of flux variability between 
the optical and near-IR imaging epochs (Table~\ref{t_info}), 
all the observations appear consistent with a spectrum which is the sum of
two-photon emission and a rather red power-law.

It is important to remember that \hi\ two-photon emission only occurs for 
electron densities $n_e \lesssim 10^4$~cm$^{-3}$.
Above that density, collisions from the $2^2S$ state of hydrogen to the $2^2P$
state lead to \lya\ emission ($2^2P$$\rightarrow$$1^2S$) instead of two-photon 
emission ($2^2S$$\rightarrow$$1^2S$).  Thus, we are not seeing emission from
gas with typical quasar broad-line region (BLR) densities
($\sim 10^{11}$~cm$^{-3}$).\footnote{It is
possible that the match of the spectral shape of \sdssj\ to that of
the \hi\ two-photon continuum is coincidental, since accretion disk
spectra are not expected to be pure power-laws \markcite{hub00}({Hubeny} {et~al.} 2000).  
If so, then we could be seeing emission from a BLR with $n_e>10^4$ cm$^{-3}$
but with unusual parameters that suppress emission from lines other than \lya.
We relegate further discussion of this rather unlikely
possibility to Appendix \ref{BLAGN}, and assume we have in fact detected
two-photon emission throughout the rest of the main text.}

The ratio of the integrated two-photon continuum flux to the \lya\ line
flux (corrected for narrow \lya\ absorption) in \sdssj\ is 
$F_{2q}/F_{Ly\alpha} = 4.60 \pm 0.66$.  The intrinsic value of this ratio is
$0.45 < F_{2q}/F_{Ly\alpha} < 0.58$ for $5,000\,{\rm K} <T<20,000\,{\rm K}$
\markcite{ost89}({Osterbrock} 1989).  Therefore, the \lya\ emission flux
in this object must have been reduced by a factor of at least $9\pm1$.  
The only plausible destruction mechanism is absorption by dust within the
\lya\ emission region, where resonant scattering of \lya\ will increase its
optical depth relative to the two-photon continuum.
(Colisional deexcitation or bound-free ionization of the $n=2$ level 
of hydrogen would affect \lya\ and two-photon emission equally.)

Assuming the UV continuum is extincted by the \lya-emitting gas, 
the maximum plausible extinction corresponds to a color excess of $E(B-V)=0.15$,
using the Small Magellanic Cloud extinction curve of \markcite{pei92}{Pei} (1992).
That extinction is the amount required to turn the \al$=-1.56$ spectral
slope of a typical SDSS quasar (\markcite{sdss73}{Vanden Berk} {et~al.} 2001)
into the estimated underlying UV spectral slope of \al$=0.15$ in \sdssj.
Extinction by $E(B-V)=0.15$ 
would reduce $F_{2q}/F_{Ly\alpha}$ by a factor of 1.9 --- the \lya\ flux would
be reduced by a factor of 2.7 and the two-photon continuum by a factor of 1.4 
in total.\footnote{Emission at wavelength $\lambda$ from a slab with total
optical depth $\tau_{\lambda}$ at that wavelength will be reduced by a factor
of $\tau_{\lambda}/(1-e^{-\tau_{\lambda}})$ from the $\tau=0$ case.}
The additional required reduction of \lya\ 
corresponds to an optical depth to scattering only 5.1 times greater in
\lya\ than in the adjacent continuum, which is easily achieved \markcite{kf98}({Korista} \& {Ferland} 1998).
However, while a reddened normal quasar spectrum plus reddened two-photon
emission is an acceptable fit to the spectrum longward of Ly$\alpha$
($\chi_{\nu}^2=1.187$) and to the $J$-band flux, it
underpredicts the
$K_s$-band flux (rest-frame 6200~\AA) by a factor of two.  \sdssj\ is too
luminous for this discrepancy to be explained by host galaxy emission, and
the relative contributions of the Balmer continuum in the observed $J$ band and
the Paschen continuum in the observed $K_s$ band will make the $J-K_s$ color
bluer, not redder, 
for all reasonable power-law indices (Fig. 4.1 of \markcite{ost89}{Osterbrock} 1989).

\section{Explaining SDSS J1136+0242} \label{WASSUP}

In this section we consider possible explanations for the unusual properties
of \sdssj.  One discounted possibility is that this object is a
gravitationally lensed Lyman-break galaxy (see Appendix \ref{LENS}).

We first summarize our conclusions regarding the spectrum of \sdssj\ so far.
A low-metallicity but dusty ($E(B-V)\lesssim0.15$) narrow-line region with 
$n_e\lesssim10^4$~cm$^{-3}$
is responsible for the weak \lya\ emission and strong two-photon
continuum which contributes $\geq$30\% of the continuum flux at 1500~\AA.
The underlying UV-optical continuum is well fit as a power-law with
$F_{\lambda} \propto \lambda^{0.15}$, much steeper than usual for AGN, though
an intrinsically bluer power-law plus reddening cannot be entirely ruled out.
The emission-line flux of \civ\ is $\lesssim$5\% of the
\lya\ flux, compared to typical values of $\sim$25\% in quasars
\markcite{sdss73}({Vanden Berk} {et~al.} 2001) and $\sim$15\% in powerful radio galaxies \markcite{deb00}({De Breuck} {et~al.} 2000).  The
\lya\ linewidth is intermediate between these two classes of objects.
The continuum is at best very weakly polarized. 
Stellar emission cannot contribute significantly to the continuum
for the same reasons the lensed Lyman-break galaxy hypothesis was rejected
(see Appendix \ref{LENS}).
We appear to be observing a featureless, power-law UV spectrum
plus \lya\ and \hi\ two-photon emission from dusty gas.

\subsection{SDSS J1136+0242 as an Unusual Narrow-Line Radio Galaxy} \label{NLAGN}

It may be reasonable to think of \sdssj\ as the spectrum of a
radio galaxy plus a featureless continuum.
The radio power of \sdssj\ is
$P_{\rm 1.4\,GHz}=10^{34.15}$
erg s$^{-1}$ Hz$^{-1}$, assuming $\alpha_{\nu}=-1.25$ near 1.4 GHz 
\markcite{deb00}({De Breuck} {et~al.} 2000). 
This radio power is within an order of magnitude of those of the powerful
$z>2$ radio galaxies studied by \markcite{deb00}{De Breuck} {et~al.} (2000).  \sdssj\ 
is radio-loud as defined by \markcite{sdss1st}{Ivezi{\' c}} {et~al.} (2002), with a logarithmic radio-optical
flux ratio $R_i=1.43$ sufficient to place it at or above the traditional
$R_i=1$ division even in the presence of reddening with $E(B-V)=0.15$.

The \lya/\civ\ flux ratio of \sdssj\ would be extreme, but not unprecedented,
for a narrow-line radio galaxy.  The
studies of \markcite{deb00,deb01}{De Breuck} {et~al.} (2000, 2001) include four objects out of $\sim$50 with 
larger \lya/\civ\ flux ratios.
Their \lya\ emission is usually spatially extended, 
but the \lya\ emission of \sdssj\ could also be extended since
SDSS spectra are obtained through 3\arcsec\ diameter fibers.
The most likely explanation for the small \civ/\lya\ ratio
in those objects, and in \sdssj\ if this model is correct,
is a low metallicity of 0.1\,$Z_{\odot}$ or less in the line-emitting gas
(Fig. 17 and \S 5.6 of \markcite{deb00}{De Breuck} {et~al.} 2000).

The \lya\ emission FWHM of \sdssj\ would also be extreme, but not unprecedented,
for a distant radio galaxy.  \sdssj\ has a FWHM larger than all but two objects 
in the sample of 52 studied by \markcite{bm00}{Baum} \& {McCarthy} (2000),  
both of which were among the 10 $z>1$ objects in the sample.
A study of $z\sim2.5$ radio galaxies by \markcite{vea97}{van Ojik} {et~al.} (1997) found \lya\ FWHMs of
670--1575 \kms, with the largest FWHM in small radio sources ($<$50 kpc).
Additionally, \lya\ absorption was seen in 9 of 10 small radio sources,
but in only 2 of 8 radio sources larger than 50 kpc.
Thus, the large \lya\ emission FWHM and presence of \lya\ absorption in
\sdssj\ is consistent with it being a compact radio galaxy.

The \lya\ luminosity of \sdssj\ is not exceptional.  \lya-emitting halos are 
often seen around powerful radio galaxies \markcite{vea97}({van Ojik} {et~al.} 1997) and radio-loud quasars
\markcite{brea92}({Bremer} {et~al.} 1992), and occasionally around radio-quiet quasars \markcite{ber99,bea03}({Bergeron} {et~al.} 1999; {Bunker} {et~al.} 2003).
The observed \lya\ luminosity of \sdssj\ is $1.18\times10^{43}$ erg s$^{-1}$,
and even the inferred intrinsic \lya\ luminosity (see \S\ref{2Q}) of
$1.06\times10^{44}$ erg s$^{-1}$ is a factor of 10 below
the largest \lya\ luminosities seen in radio galaxies \markcite{vm03}({Villar-Mart{\'{\i}}n} {et~al.} 2003).

The luminosity of the two-photon continuum is nearly unprecedented, however.
If only the two-photon continuum and \lya\ were observed,
\sdssj\ would still have $r=21.5$, a magnitude equalled by fewer than a handful
of powerful radio galaxies at $2\leq z\leq 3$ \markcite{cmv88,pen97,deb01}({Chambers}, {Miley}, \& {van  Breugel} 1988; {Pentericci} {et~al.} 1997; {De Breuck} {et~al.} 2001).
The $K_s$ magnitude of \sdssj\ ties with the brightest observed among
powerful radio galaxies at $2\leq z\leq 3$ in the sample of \markcite{wrjb03}{Willott} {et~al.} (2003)
and is $\simeq$1\fm25 brighter than their best-fit $K-z$ relation.
Two-photon continuum is not usually considered to be the dominant explanation
for the UV continua of powerful radio galaxies.  However, it clearly contributes
a similar fraction of the continuum luminosity in some radio galaxies as it does
in \sdssj\ \markcite{ver01}({Vernet} {et~al.} 2001), and the fact that dust mixed with the emitting gas can
modify the characteristic shape of the two-photon continuum may mean that it is
more common than has been concluded from analyses that neglect that effect.

The volume $V$ of the region emitting the \lya\ line and the two-photon
continuum can be estimated using the relation
$L_{Ly\alpha} = 4 \times 10^{-24} n_e^2 f V$ erg s$^{-1}$ \markcite{vm03}({Villar-Mart{\'{\i}}n} {et~al.} 2003)
with the constraint that $n_e \leq 10^4$ cm$^{-3}$.  Relevant estimates of the
filling factor of the \lya-emitting gas, $f$, range from 10$^{-4}$ to $10^{-8}$
(\S\,4.2.4 of \markcite{vm03}{Villar-Mart{\'{\i}}n} {et~al.} 2003), but if we 
apply the above equation to the intrinsic $L_{Ly\alpha}$
inferred from the SDSS spectroscopy then $f$ must be such that the emission
region in \sdssj\ is smaller in angular size than the SDSS fiber diameter of
3\arcsec\ (corresponding to 11.8 kpc at $z=2.4917$).  Assuming a typical radio
galaxy narrow-line region (NLR) with $n_e=10^2$ cm$^{-3}$ \markcite{vm03}({Villar-Mart{\'{\i}}n} {et~al.} 2003), a
relatively high filling factor of $f>2.5\times10^{-5}$ is inferred.  
Spatially resolved observations of the \lya\ emission
would clearly be useful to clarify its origin and to
jointly constrain the density and filling factor of the \lya-emitting gas.

One important difference between \sdssj\ and the brightest $z\simeq2.5$ radio
galaxies is that at least some of the latter objects have much higher
polarizations (5-10\% vs. $<$1\% for \sdssj; \markcite{ver01}{Vernet} {et~al.} 2001).  
This difference can be explained if \sdssj\ is viewed 
close to face-on, unlike most powerful radio galaxies.  
Such a viewing angle maximizes the symmetry
of any scattering, thereby decreasing the observed polarization.  For a
biconical scattering region geometry with opening half-angle 45\arcdeg\ and
a maximum $P=17$\% at a viewing angle from the cone axis of $\vartheta=90\arcdeg$, 
an observed $P<1$\% is possible at $\vartheta<8\arcdeg$ \markcite{mdsa96}({Manzini} \& {di Serego Alighieri} 1996).
The larger the opening angle of the bicones, the larger the range of $\vartheta$
for which $P<1$\%, and of course $P=0$\% for a spherical scattering region.

\subsection{Possible Origins of the Featureless UV Continuum} \label{CONT}

The scenario outlined above for \sdssj\ --- a face-on radio galaxy with an
extended, dusty, low-metallicity, \lya-emitting NLR ---
explains many of its characteristics.
We now discuss possible explanations for 
why we see a luminous, featureless, and variable ultraviolet continuum but no
emission from the broad-line region,
properties which are not immediately explained by this model.
\begin{itemize}
\item The observed variability, among other arguments, rules out 
a stellar origin for the observed UV continuum (see Appendix \ref{LENS}). 
\item If by chance a gap in the obscuring matter enables us to see the continuum
source in this object, we should also see emission from the broad-line region
unless the gap is very small (since the continuum source is smaller than the 
BLR) or the BLR produces unusually weak lines (see Appendix \ref{BLAGN}).  
Some broad absorption line (BAL) quasars have very weak 
metal emission lines, but no BAL outflow has ever been seen to
absorb emission lines without also strongly absorbing the continuum.
\item A related possibility is that we {\em are} seeing the continuum
source and the BLR, but that the latter has a very high density
of $n_H \gtrsim 10^{13}$ cm$^{-3}$ and has thermalized
so that it emits only blackbody radiation. 
The dominant range of temperatures expected for such emission is
$6000\lesssim T\lesssim 10000$~K \markcite{fer99}({Ferland} 1999), and our
photometry covers the peak
wavelengths of such blackbodies.
There is no clear sign of such emission in the photometry or spectroscopy.
It is true that blackbody emission could help explain the deficit of $K_s$-band
flux found when the spectrum is modeled as a reddened power-law with \al$=-1.56$
(\S\ref{2Q}), but the temperature of such putative emission would have to be
$4000~{\rm K}\lesssim T\lesssim 5500~{\rm K}$.  This is too narrow a 
temperature range, and too far below the expected temperatures,
to be considered a detection of thermalized BLR emission.
\item Perhaps we have an unobscured sightline to a central engine which simply
lacks a BLR.  For example, \markcite{nmm03}{Nicastro}, {Martocchia}, \& {Matt} (2003) have suggested that AGN with accretion
rates $<10^{-3}$ of the Eddington rate might lack broad lines.  But if \sdssj\ 
was accreting at such a low rate, explaining its observed luminosity would 
require an implausibly large black hole mass of $\gtrsim10^{11.4}~M_{\odot}$
\markcite{pet97,md04}({Peterson} 1997; {McLure} \& {Dunlop} 2004).
Thus, a different explanation than that of \markcite{nmm03}{Nicastro} {et~al.} (2003) would need to be found
if \sdssj\ is shown to lack a BLR.
\item Seyfert 2 galaxies are known to have an unpolarized, featureless UV
continuum which is not attributable to starlight, but 
is plausibly due to optically thin free-free emission \markcite{tran95}({Tran} 1995).
However, this emission component 
usually arises from outside the broad-line region, where any variability
timescale should be longer than the rest-frame 40-70 days observed in \sdssj.
Moreover, this emission
component typically accounts for only $\sim$4\% of the total UV flux of a
Seyfert 1 galaxy, whereas it would have to account
for $\sim$70\% of the UV flux in \sdssj.
\item The continuum could be relativistically beamed synchrotron emission from
a jet pointing nearly along our line of sight.
The beaming factor must be at least $\sim$15
to explain the low equivalent widths of \civ\ and \ciii. 
Beaming would be consistent with the face-on orientation of \sdssj\ invoked to
explain the absence of highly polarized scattered light.
While synchrotron-dominated objects (blazars) usually have high intrinsic
polarizations, there are at least two possible explanations for why that 
is not the case here.  One is that such polarization is highly variable: some 
BL~Lac objects with $P>3$\% at one epoch have been seen to have $P<1$\% at 
another epoch \markcite{jse93,vw98}({Jannuzi}, {Smith}, \& {Elston} 1993; {Visvanathan} \& {Wills} 1998).  The other is that the synchrotron emission
in blazars typically arises from several emission regions, and that unless 
those regions have aligned polarization vectors the overall polarization will
be reduced (\S\,4.1.1 of \markcite{ls00}{Lister} \& {Smith} 2000).
However, there are two strong arguments against this beaming hypothesis.
First, the spectral energy distribution (SED) of \sdssj\ is much steeper
between radio and UV frequencies than is the case for blazars 
(Fig. 12 of \markcite{fea98}{Fossati} {et~al.} 1998).
Second, blazars are typically polarized parallel to the radio jet axis, because 
the synchrotron emission arises in shocks where compression yields
magnetic fields oriented perpendicular to the jet flow
(recall that the polarization and magnetic field vectors are perpendicular).
But in \sdssj, if we identify the radio major axis (\S\ref{RADIO}) as the jet
axis then the polarization is consistent with being perpendicular to it
($\Delta\theta=81\pm20$\arcdeg).
\item
\sdssj\ may be a higher-redshift, lower-luminosity analog of the quasar 
\pg\ \markcite{mcd95}({McDowell} {et~al.} 1995),
which has very weak, very broad, highly blueshifted high-ionization emission
lines (\civ\ has REW=4.6$\pm$2~\AA\ and is blueshifted by 
10200$\pm$1200\,\kms\ from \ha; both lines have a FWHM of $\simeq$7000\,\kms).
The SED of \pg\ is a very good match to the median radio-quiet 
quasar SED of \markcite{elv94}{Elvis} {et~al.} (1994), and so its weak lines cannot be explained by a very
soft ionizing continuum \markcite{lhj04}({Leighly}, {Halpern}, \& {Jenkins} 2004) or a relativistically beamed continuum.
Nonetheless, \markcite{bbb03}{Blundell}, {Beasley}, \&  {Bicknell} (2003) have presented strong evidence for the existence of
a relativistic jet in \pg\ oriented very close to our line of sight.
Unlike most blazars, \pg\ has a very low polarization of $P=0.24\pm0.16$\%
at a position angle
roughly perpendicular to the jet axis ($\Delta\theta=70\pm20$\arcdeg, assuming
the jet axis is the A-B axis in \markcite{bbb03}{Blundell} {et~al.} 2003).
These observations
place \pg\ in the class of low-polarization radio quasars \markcite{ls00}({Lister} \& {Smith} 2000).
Such objects
are thought to represent quiescent blazars whose jets do not
harbor strong shocks, resulting in 
magnetic fields which are not necessarily oriented perpendicular to the flow.

\pg\ and \sdssj\ are similar in that they have weak UV emission lines, low
polarizations apparently oriented perpendicular to the radio jet axis,
and nearly face-on orientations.  It is not known why \pg\ has such
unusual emission-line properties,\footnote{
A small number of `lineless' quasars have been identified 
\markcite{fan99b,and01,lhj04}({Fan} {et~al.} 1999; {Anderson} {et~al.} 2001; {Leighly} {et~al.} 2004)
which have weak emission lines for reasons apparently unrelated to relativistic
beaming, but which may be extreme examples of the Baldwin effect.
However, \sdssj\ is not luminous enough for its weak line emission to be
plausibly explained by the Baldwin effect, even accounting for the large 
scatter in that relation --- \sdssj\ is 2.5 times less luminous than \pg\ and 
yet has a \civ\ REW two times smaller.  
}
but whatever the explanation is, it may apply to both objects.
If \sdssj\ is similar to \pg, we expect that optical spectra with a higher SNR
will reveal very broad, weak, highly blueshifted emission
lines and that near-IR spectra will reveal strong \feii\ emission \markcite{mcd95}({McDowell} {et~al.} 1995).
In addition,
as a low-polarization radio quasar, \sdssj\ should have a radio core 
with the same, low polarization as the optical continuum and with the same
polarization position angle roughly perpendicular to the jet axis.
\end{itemize}

\section{Conclusion} \label{CON}

The newly discovered, radio-loud AGN \sdssj\ is unusual in possessing 
a strong continuum without accompanying strong metal-line emission.
Of $\sim$2400 quasars in the SDSS Data Release 1 quasar catalog \markcite{dr1q}({Schneider} {et~al.} 2003)
with $z\geq2.12$, sufficient to place \lya\ within the
spectral coverage of the SDSS spectrographs, it is the only object with strong,
relatively narrow \lya\ emission and no accompanying metal-line emission
(excepting broad absorption line quasars where such emission is absorbed
along with part or all of the continuum).

The spectrum of \sdssj\ can generally be understood as that of a very optically
luminous radio galaxy, in terms of \lya\ emission flux, velocity width, 
associated narrow \lya\ and \civ\ absorption, \civ/\lya\ ratio (from a
low-metallicity NLR), and even two-photon continuum strength 
(\S\ref{2Q} and \S\ref{NLAGN}).

However, \sdssj\ has a very strong UV continuum not seen in radio galaxies
(\S\ref{CONT}).  This continuum is unpolarized,
suggesting a face-on orientation for \sdssj.
The continuum could be relativistically beamed synchrotron emission 
which swamps the intrinsic broad-line emission, but the SED
of \sdssj\ is not a good match to known synchrotron-dominated objects.
A more likely possibility is that \sdssj\ is a \pg\ analog: a low-polarization
radio quasar which for some unknown reason has
very weak, very broad, highly blueshifted high-ionization emission lines.
A better optical spectrum is needed to determine if such features are present
in \sdssj.
This scenario predicts polarization perpendicular to the axis of the radio
jet, which is tentatively observed to be the case in both \sdssj\ and \pg\ 
but requires confirmation with better polarization data.

In the unlikely event the \lya\ 
is from a broad-line region of typical density, the dip in the continuum 
otherwise attributed to \hi\ two-photon emission is unexplained.
Also, the BLR would have to
consist predominantly of low-ionization or low-metallicity gas
(Appendix \ref{BLAGN}).
If the former, it is unclear what about \sdssj\ causes its lack of
the high-ionization gas which dominates the broad-line regions of most AGNs.
If the latter, metallicities $\lesssim$1\% of solar are required.  

Further investigation of \sdssj\ would benefit from 
better optical spectra (preferably spatially resolved to search for and
measure any spatial extent of \lya),
better optical polarimetry (or at least another epoch of the same quality),
deep, high-resolution radio imaging and polarimetry to determine the jet axis,
near-IR spectroscopy to study the rest-frame optical spectrum,
and multiwavelength photometry to determine if its spectral energy distribution
can be reconciled with those of blazars.

\acknowledgements
We thank D. Johnston, M. Brotherton, M. De Robertis, J. Krolik, G. Richards,
C. Ryan, S. Savaglio, M. SubbaRao, P. Wiita and J. Willis for comments and 
assistance.
PBH acknowledges support from Fundaci\'{o}n Andes and the Department of
Astrophysical Sciences at Princeton University,
EH \& KG
from the David and Lucille Packard Foundation,
DPS from NSF grant AST03-07582,
and MNO from the REU program via a supplement to NSF grant AST-0071091
to Princeton University.
Funding for the creation and distribution of the SDSS Archive has been provided
by the Alfred P. Sloan Foundation, the Participating Institutions, the National
Aeronautics and Space Administration, the National Science Foundation, the U.S.
Department of Energy, the Japanese Monbukagakusho, and the Max Planck Society.
The SDSS Web site is http://www.sdss.org/.  
The SDSS is managed by the Astrophysical Research Consortium (ARC) for the 
Participating Institutions:
The University of Chicago, Fermilab, the Institute for Advanced Study, the 
Japan Participation Group, The Johns Hopkins University, Los Alamos National 
Laboratory, the Max-Planck-Institute for Astronomy (MPIA), the 
Max-Planck-Institute for Astrophysics (MPA), New Mexico State University, 
University of Pittsburgh, Princeton University, the United States Naval 
Observatory, and the University of Washington.

\begin{appendix}

\section{An Unusual Broad-Line AGN?} \label{BLAGN}

If the match of the spectral shape of \sdssj\ to that
of the \hi\ two-photon continuum is coincidental, 
we might be seeing emission from a BLR with $n_e>10^4$ cm$^{-3}$
and unusual parameters that suppress emission from lines other than \lya.

\begin{figure}
\epsscale{1.00}
\plotone{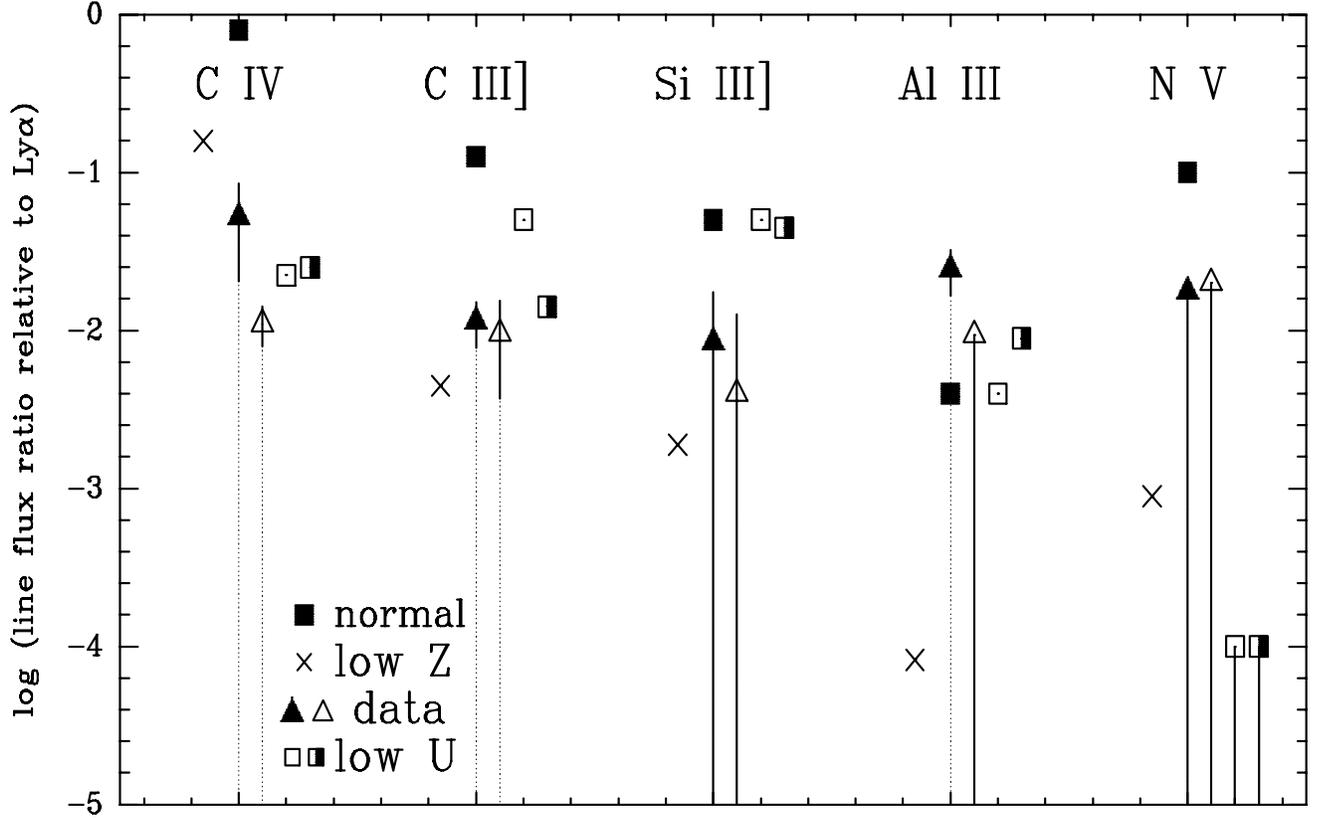}
\caption[]{
\singlespace 
Line ratios for SDSS J1136+0242 and various photoionization simulations.
Solid triangles (with $1 \sigma$ error bars) are the variable-$\lambda_{cen}$
measurements from Table \ref{t_ratio},
open triangles and error bars are fixed-$\lambda_{cen}$ measurements,
filled squares are from the baseline photoionization model of F99,
crosses are from the low metallicity simulation ($\log Z/Z_{\odot}=-2$),
open squares are from the low ionization parameter simulation ($\log U = -3$),
and half-filled squares are estimates of a low ionization parameter simulation
with a density $\log n_H = 11$ instead of the baseline $\log n_H = 10$.
Solid lines extending off the bottom of the plot are limits consistent with zero.
The dotted lines are reminders that the neither of the fits to the data allow
the continuum to vary simultaneously with the lines, and thus all measurements
may be upper limits.
}\label{f_rats}
\end{figure}

\markcite{fer99}{Ferland} (1999), hereafter F99, discuss the line emission from a single quasar
broad-line region gas cloud as a function of 
many different physical parameters.  
\civ\ emission is an important coolant in photoionized gas, and so weak \civ\ 
emission indicates gas with low metallicity ($\log Z/Z_{\odot}\leq -1$),
low ionization ($\log U \leq -2$), or both.
Tentative emission line flux ratios and limits for \sdssj\ are listed in 
Table \ref{t_ratio}, along with ratios from two photoionization simulations 
given in F99.
Figure \ref{f_rats} compares several sets of line ratios:
solid triangles (with $1\sigma$ error bars) are the variable-$\lambda_{cen}$
measurements,
open triangles are the fixed-$\lambda_{cen}$ measurements,
crosses are from a low metallicity simulation ($\log Z/Z_{\odot}=-2$),
open squares are from a low ionization parameter simulation
($\log U = -3$, where $U$ is the ratio of the photon
density at the cloud's illuminated face to the hydrogen nucleus density $n_H$),
and half-filled squares are estimates of a low ionization parameter
simulation with a density $n_H = 10^{11}$ cm$^{-3}$
instead of the baseline $n_H = 10^{10}$ cm$^{-3}$
(see below).

The low-metallicity model is reasonably consistent with the data,
but would be strongly rejected if the nominal detection of \aliii\ in the
variable-$\lambda_{cen}$ fit is correct.
(The fixed-$\lambda_{cen}$ fit yields only an upper limit on \aliii.)
Note that the presence of associated \civ\ absorption with saturated line ratios
does {\em not} rule out the low-metallicity model.  Even low metallicity
\civ\ absorption systems are often saturated at the SDSS spectral resolution,
and the associated absorption systems often seen in AGN can be located in
nearby galaxies or in host galaxy disks or halos, as well as 
near the nucleus \markcite{hf99}({Hamann} \& {Ferland} 1999).

Of the models plotted, the line ratios of \sdssj\ are most consistent with 
emission from low-ionization,
high-density clouds.  The greatest discrepancy (2$-$3$\sigma$) in that scenario
is that \Siiii\ is predicted to be $\sim$10 times stronger than observed.
Nonetheless, it is preferred over the normal density low-ionization simulation
because the latter predicts too strong a \ciii\ line, at the 4$-$6$\sigma$ 
level.  

If we adopt the locally optimally-emitting cloud model pioneered by 
\markcite{loc95}{Baldwin} {et~al.} (1995), in which the emitting gas is assumed to span a wide range in 
density and distance (and therefore ionization parameter), a low metallicity
scenario is preferable.  Otherwise, we would have to explain why gas exists
close to the black hole (to produce the UV continuum source) and far away
from it (to produce low-ionization emission), but not at intermediate
distances where broad metal-line emission is normally produced.

\section{A Lensed Lyman-break Galaxy?} \label{LENS}

We were initially enamored of the possibility that \sdssj\ could be a
lensed Lyman-break galaxy (LBG) similar to MS1512-cB58 \markcite{yea96}({Yee} {et~al.} 1996), 
but with \lya\ in emission.  This was due in part to its spectrum
being radically different from a typical broad-line AGN and in part
to the dip in the spectrum between \lya\ and $\sim$1280~\AA, which
is not generally seen in quasars but is seen in LBGs \markcite{sha02}({Shapley} {et~al.} 2002).
However, the evidence against lensing is overwhelming.

First, \sdssj\ is photometrically variable at the 99.8\% confidence level,
with $\Delta m_{AB} = 0.08\pm0.02$ (averaged over five bands)
over 67 rest-frame days.  Such variability is common in AGN but would be
inexplicable for a Lyman-break galaxy.

Second, \sdssj\ is unresolved in the SDSS $ugriz$ imaging and in $J$ imaging
obtained on UT 2003 April 19 at the Canada-France-Hawaii Telescope (CFHT)
with the CFHTIR camera. 
In the CFHT image, \sdssj\ has FWHM=0\farcs59, compared to
FWHM=$0\farcs60\pm0\farcs01$ measured from 6 stars in the same image.
A double-image lens with separation 0\farcs2 would have a FWHM=0\farcs63
in 0\farcs60 seeing, and so lensing with separation $\geq 0\farcs2$ 
is ruled out at the $\sim 3-4 \sigma$ level.
This does not necessarily rule out a high magnification lens ---
for example, APM 08279+5255 has a separation of 0\farcs378 
and a magnification of $\sim$40$-$90 \markcite{iea99}({Ibata} {et~al.} 1999) --- but it does
considerably narrow the parameter space accessible to such a lens.
Similarly, a cluster-lens such as MS1512-cB58 is rendered unlikely by the 
non-detection of any other object within 11\arcsec\ in either the CTIO,
CFHT or SDSS imaging, to limits $\sim$3 magnitudes fainter than \sdssj\ in
$i$, $\sim$2 magnitudes in $J$, and $\sim$1 magnitude in $K_s$.
A lensing cluster would have to have a redshift $z\gtrsim$1 
to avoid detection in those data.

Third, the spectral properties of \sdssj\ are very different from those of LBGs.
Its rest-frame UV spectral index ($\alpha_{\lambda}=0.15$) is
much redder than is typical for LBGs ($-1.09 <$\al$< -0.73$; \markcite{sha02}{Shapley} {et~al.} 2002).
And if it was a LBG, \sdssj\ would go against the trend found by \markcite{sha02}{Shapley} {et~al.} (2002)
for the UV continuum to become bluer with increasing \lya\ emission strength.
Additionally, perusal of model UV spectra for a wide range of young stellar
population models \markcite{lea99}({Leitherer} {et~al.} 1999) 
reveals none with continua as featureless as that of \sdssj.

We therefore conclude that \sdssj\ is not a lensed Lyman-break galaxy.

\end{appendix}

\bibliography{}

\begin{thebibliography}{}

\bibitem[{Aller} 1984]{all84}
{Aller}, L.~H., ed. 1984, {Physics of Thermal Gaseous Nebulae}

\bibitem[{Anderson}, {Fan}, {Richards}, {Schneider},  {Strauss}, {Vanden Berk}, {Gunn}, {Knapp}, {Schlegel}, {Voges}, {Yanny},  {Bahcall}, {Bernardi}, {Brinkmann}, {Brunner}, {Csab{\' a}i}, {Doi},  {Fukugita}, {Hennessy}, {Ivezi{\' c}}, {Kunszt}, {Lamb}, {Loveday}, {Lupton},  {McKay}, {Munn}, {Nichol}, {Szokoly}, \& {York} 2001]{and01}
{Anderson}, S.~F., {Fan}, X., {Richards}, G.~T., {Schneider}, D.~P., {Strauss},  M.~A., {Vanden Berk}, D.~E., {Gunn}, J.~E., {Knapp}, G.~R., {et al.}, 2001, \aj, 122, 503

\bibitem[{Baldwin}, {Ferland}, {Korista}, \&  {Verner} 1995]{loc95}
{Baldwin}, J., {Ferland}, G., {Korista}, K., \& {Verner}, D. 1995, \apjl, 455,  L119

\bibitem[{Baum} \& {McCarthy} 2000]{bm00}
{Baum}, S.~A. \& {McCarthy}, P.~J. 2000, \aj, 119, 2634

\bibitem[{Becker}, {White}, \& {Helfand} 1995]{bwh95}
{Becker}, R.~H., {White}, R.~L., \& {Helfand}, D.~J. 1995, \apj, 450, 559

\bibitem[{Bergeron}, {Petitjean}, {Cristiani},  {Arnouts}, {Bresolin}, \& {Fasano} 1999]{ber99}
{Bergeron}, J., {Petitjean}, P., {Cristiani}, S., {Arnouts}, S., {Bresolin},  F., \& {Fasano}, G. 1999, \aap, 343, L40

\bibitem[{Blanton}, {Lin}, {Lupton}, {Maley}, {Young},  {Zehavi}, \& {Loveday} 2003]{sdss89}
{Blanton}, M.~R., {Lin}, H., {Lupton}, R.~H., {Maley}, F.~M., {Young}, N.,  {Zehavi}, I., \& {Loveday}, J. 2003, \aj, 125, 2276

\bibitem[{Blundell}, {Beasley}, \&  {Bicknell} 2003]{bbb03}
{Blundell}, K.~M., {Beasley}, A.~J., \& {Bicknell}, G.~V. 2003, \apjl, 591,  L103

\bibitem[{Bremer}, {Fabian}, {Sargent}, {Steidel},  {Boksenberg}, \& {Johnstone} 1992]{brea92}
{Bremer}, M.~N., {Fabian}, A.~C., {Sargent}, W.~L.~W., {Steidel}, C.~C.,  {Boksenberg}, A., \& {Johnstone}, R.~M. 1992, \mnras, 258, 23P

\bibitem[{Bunker}, {Smith}, {Spinrad}, {Stern}, \&  {Warren} 2003]{bea03}
{Bunker}, A., {Smith}, J., {Spinrad}, H., {Stern}, D., \& {Warren}, S. 2003,  \apss, 284, 357

\bibitem[{Chambers}, {Miley}, \& {van  Breugel} 1988]{cmv88}
{Chambers}, K.~C., {Miley}, G.~K., \& {van Breugel}, W.~J.~M. 1988, \apjl, 327,  L47

\bibitem[{De Breuck}, {R{\" o}ttgering}, {Miley},  {van Breugel}, \& {Best} 2000]{deb00}
{De Breuck}, C., {R{\" o}ttgering}, H., {Miley}, G., {van Breugel}, W., \&  {Best}, P. 2000, \aap, 362, 519

\bibitem[{De Breuck}, {van Breugel}, {R{\"  o}ttgering}, {Stern}, {Miley}, {de Vries}, {Stanford}, {Kurk}, \&  {Overzier} 2001]{deb01}
{De Breuck}, C., {van Breugel}, W., {R{\" o}ttgering}, H., {Stern}, D.,  {Miley}, G., {de Vries}, W., {Stanford}, S.~A., {Kurk}, J., {et al.}, 2001, \aj, 121, 1241

\bibitem[{Elvis}, {Wilkes}, {McDowell}, {Green},  {Bechtold}, {Willner}, {Oey}, {Polomski}, \& {Cutri} 1994]{elv94}
{Elvis}, M., {Wilkes}, B.~J., {McDowell}, J.~C., {Green}, R.~F., {Bechtold},  J., {Willner}, S.~P., {Oey}, M.~S., {Polomski}, E., {et al.}, 1994,  \apjs, 95, 1

\bibitem[{Fan}, {Strauss}, {Gunn}, {Lupton}, {Carilli},  {Rupen}, {Schmidt}, {Moustakas}, {Davis}, {Annis}, {Bahcall}, {Brinkmann},  {Brunner}, {Csabai}, {Doi}, {Fukugita}, {Heckman}, {Hennessy}, {Hindsley},  {Ivezi{\' c} }, {Knapp}, {Lamb}, {Munn}, {Pauls}, {Pier}, {Rockosi},  {Schneider}, {Szalay}, {Tucker}, \& {York} 1999]{fan99b}
{Fan}, X., {Strauss}, M.~A., {Gunn}, J.~E., {Lupton}, R.~H., {Carilli}, C.~L.,  {Rupen}, M.~P., {Schmidt}, G.~D., {Moustakas}, L.~A., {et al.}, 1999, \apjl, 526, L57

\bibitem[{Ferland} 1999]{fer99}
{Ferland}, G. 1999, in Quasars and Cosmology, eds. G. Ferland \& J. Baldwin,  (San Francisco: ASP), 147

\bibitem[{Fossati}, {Maraschi}, {Celotti}, {Comastri},  \& {Ghisellini} 1998]{fea98}
{Fossati}, G., {Maraschi}, L., {Celotti}, A., {Comastri}, A., \& {Ghisellini},  G. 1998, \mnras, 299, 433

\bibitem[{Fukugita}, {Ichikawa}, {Gunn}, {Doi},  {Shimasaku}, \& {Schneider} 1996]{fuk96}
{Fukugita}, M., {Ichikawa}, T., {Gunn}, J.~E., {Doi}, M., {Shimasaku}, K., \&  {Schneider}, D.~P. 1996, \aj, 111, 1748

\bibitem[{Gunn}, {Carr}, {Rockosi}, {Sekiguchi}, {Berry},  {Elms}, {de Haas}, {Ivezi{\'c} }, {Knapp}, {Lupton}, {Pauls}, {Simcoe},  {Hirsch}, {Sanford}, {Wang}, {York}, {Harris}, {Annis}, {Bartozek},  {Boroski}, {Bakken}, {Haldeman}, {Kent}, {Holm}, {Holmgren}, {Petravick},  {Prosapio}, {Rechenmacher}, {Doi}, {Fukugita}, {Shimasaku}, {Okada}, {Hull},  {Siegmund}, {Mannery}, {Blouke}, {Heidtman}, {Schneider}, {Lucinio}, \&  {Brinkman} 1998]{gun98}
{Gunn}, J.~E., {Carr}, M., {Rockosi}, C., {Sekiguchi}, M., {Berry}, K., {Elms},  B., {de Haas}, E., {Ivezi{\'c} }, {\v Z}., {et al.}, 1998, \aj,  116, 3040

\bibitem[{Hamann} \& {Ferland} 1999]{hf99}
{Hamann}, F. \& {Ferland}, G. 1999, \araa, 37, 487

\bibitem[{Hamann}, {Korista}, {Ferland}, {Warner}, \&  {Baldwin} 2002]{hea02}
{Hamann}, F., {Korista}, K.~T., {Ferland}, G.~J., {Warner}, C., \& {Baldwin},  J. 2002, \apj, 564, 592

\bibitem[{Hogg}, {Finkbeiner}, {Schlegel}, \&  {Gunn} 2001]{sdss82}
{Hogg}, D., {Finkbeiner}, D., {Schlegel}, D., \& {Gunn}, J. 2001, \aj, 122,  2129

\bibitem[{Hubeny}, {Agol}, {Blaes}, \&  {Krolik} 2000]{hub00}
{Hubeny}, I., {Agol}, E., {Blaes}, O., \& {Krolik}, J.~H. 2000, \apj, 533, 710

\bibitem[{Ibata}, {Lewis}, {Irwin}, {Leh{\' a}r}, \&  {Totten} 1999]{iea99}
{Ibata}, R.~A., {Lewis}, G.~F., {Irwin}, M.~J., {Leh{\' a}r}, J., \& {Totten},  E.~J. 1999, \aj, 118, 1922

\bibitem[{Ivezi{\' c}}, {Menou}, {Knapp},  {Strauss}, {Lupton}, {Vanden Berk}, {Richards}, {Tremonti}, {Weinstein},  {Anderson}, {Bahcall}, {Becker}, {Bernardi}, {Blanton}, {Eisenstein}, {Fan},  {Finkbeiner}, {Finlator}, {Frieman}, {Gunn}, {Hall}, {Kim}, {Kinkhabwala},  {Narayanan}, {Rockosi}, {Schlegel}, {Schneider}, {Strateva}, {SubbaRao},  {Thakar}, {Voges}, {White}, {Yanny}, {Brinkmann}, {Doi}, {Fukugita},  {Hennessy}, {Munn}, {Nichol}, \& {York} 2002]{sdss1st}
{Ivezi{\' c}}, {\v Z}., {Menou}, K., {Knapp}, G.~R., {Strauss}, M.~A.,  {Lupton}, R.~H., {Vanden Berk}, D.~E., {Richards}, G.~T., {Tremonti}, C., {et al.}, 2002,  \aj, 124, 2364

\bibitem[{Jannuzi}, {Smith}, \& {Elston} 1993]{jse93}
{Jannuzi}, B.~T., {Smith}, P.~S., \& {Elston}, R. 1993, \apjs, 85, 265

\bibitem[{Korista} \& {Ferland} 1998]{kf98}
{Korista}, K. \& {Ferland}, G. 1998, \apj, 495, 672

\bibitem[{Leighly}, {Halpern}, \& {Jenkins} 2004]{lhj04}
{Leighly}, K.~M., {Halpern}, J.~P., \& {Jenkins}, E.~P. 2004, to appear in AGN  Physics with the Sloan Digital Sky Survey, eds. G. T. Richards \& P. B. Hall, in press (astro-ph/0402535)

\bibitem[{Leitherer}, {Schaerer}, {Goldader},  {Delgado}, {Robert}, {Kune}, {de Mello}, {Devost}, \& {Heckman} 1999]{lea99}
{Leitherer}, C., {Schaerer}, D., {Goldader}, J.~D., {Delgado}, R.~M.~G.,  {Robert}, C., {Kune}, D.~F., {de Mello}, D.~F., {Devost}, D., {et al.}, 1999, \apjs, 123, 3

\bibitem[{Lister} \& {Smith} 2000]{ls00}
{Lister}, M.~L. \& {Smith}, P.~S. 2000, \apj, 541, 66

\bibitem[{Lupton}, {Gunn}, \& {Szalay} 1999]{sdss26}
{Lupton}, R.~H., {Gunn}, J.~E., \& {Szalay}, A.~S. 1999, \aj, 118, 1406

\bibitem[{Manzini} \& {di Serego Alighieri} 1996]{mdsa96}
{Manzini}, A. \& {di Serego Alighieri}, S. 1996, \aap, 311, 79

\bibitem[{McDowell}, {Canizares}, {Elvis},  {Lawrence}, {Markoff}, {Mathur}, \& {Wilkes} 1995]{mcd95}
{McDowell}, J.~C., {Canizares}, C., {Elvis}, M., {Lawrence}, A., {Markoff}, S.,  {Mathur}, S., \& {Wilkes}, B.~J. 1995, \apj, 450, 585

\bibitem[{McLure} \& {Dunlop} 2004]{md04}
{McLure}, R.~J. \& {Dunlop}, R.~J. 2004, \mnras, in press (astro-ph/0310267)

\bibitem[{Morton}, {York}, \& {Jenkins} 1988]{myj88}
{Morton}, D.~C., {York}, D.~G., \& {Jenkins}, E.~B. 1988, \apjs, 68, 449

\bibitem[{Nicastro}, {Martocchia}, \& {Matt} 2003]{nmm03}
{Nicastro}, F., {Martocchia}, A., \& {Matt}, G. 2003, \apjl, 589, L13

\bibitem[{Osterbrock} 1989]{ost89}
{Osterbrock}, D.~E. 1989, Astrophysics of Gaseous Nebulae and Active Galactic  Nuclei (Mill Valley: University Science Books)

\bibitem[{Pei} 1992]{pei92}
{Pei}, Y.~C. 1992, \apj, 395, 130

\bibitem[{Pentericci}, {Roettgering}, {Miley},  {Carilli}, \& {McCarthy} 1997]{pen97}
{Pentericci}, L., {Roettgering}, H.~J.~A., {Miley}, G.~K., {Carilli}, C.~L., \&  {McCarthy}, P. 1997, \aap, 326, 580

\bibitem[{Peterson} 1997]{pet97}
{Peterson}, B.~M. 1997, Active Galactic Nuclei (Cambridge: Cambridge University  Press)

\bibitem[{Pier}, {Munn}, {Hindsley}, {Hennessy}, {Kent},  {Lupton}, \& {Ivezi{\' c}} 2003]{sdss153}
{Pier}, J.~R., {Munn}, J.~A., {Hindsley}, R.~B., {Hennessy}, G.~S., {Kent},  S.~M., {Lupton}, R.~H., \& {Ivezi{\' c}}, {\v Z}. 2003, \aj, 125, 1559

\bibitem[{Probst}, {Montane}, {Warner}, {Boccas},  {Bonati}, {Galvez}, {Tighe}, {Ashe}, {van der Bliek}, \& {Blum} 2003]{ispi}
{Probst}, R.~G., {Montane}, A., {Warner}, M., {Boccas}, M., {Bonati}, M.,  {Galvez}, R., {Tighe}, R., {Ashe}, M.~C., {et al.}, 2003, in Instrument Design and Performance for Optical/Infrared  Ground-based Telescopes, eds. M. Iye \& A. F. M. Moorwood, Proceedings of the  SPIE, 4841, 411--419

\bibitem[{Richards}, {Fan}, {Newberg},  {Strauss}, {Vanden Berk}, {Schneider}, {Yanny}, {Boucher}, \&  {Burles} 2002a]{sdssqtarget}
{Richards}, G.~T., {Fan}, X., {Newberg}, H.~J., {Strauss}, M.~A., {Vanden  Berk}, D.~E., {Schneider}, D.~P., {Yanny}, B., {Boucher}, A., {et al.}, 2002a, \aj, 123, 2945

\bibitem[{Richards}, {Vanden Berk},  {Reichard}, {Hall}, {Schneider}, {SubbaRao}, {Thakar}, \& {York} 2002b]{emshift}
{Richards}, G.~T., {Vanden Berk}, D.~E., {Reichard}, T.~A., {Hall}, P.~B.,  {Schneider}, D.~P., {SubbaRao}, M., {Thakar}, A.~R., \& {York}, D.~G.  2002b, \aj, 124, 1

\bibitem[{{Richards} {et~al.}(2003){Richards}, {Hall}, {Vanden Berk}, {Strauss}, {Schneider}, {Weinstein}, {Reichard}, {York}, {Knapp}, {Fan}, {Ivezi{\' c}}, {Brinkmann}, {Budav{\' a}ri}, {Csabai}, \& {Nichol}}]{redqso}
{Richards}, G.~T., {Hall}, P.~B., {Vanden Berk}, D.~E., {Strauss}, M.~A., {Schneider}, D.~P., {Weinstein}, M.~A., {Reichard}, T.~A., {York}, D.~G., {Knapp}, G.~R., {Fan}, X., {Ivezi{\' c}}, {\v Z}., {Brinkmann}, J., {Budav{\' a}ri}, T., {Csabai}, I., \& {Nichol}, R.~C. 2003, \aj, 126, 1131

\bibitem[{Schlegel}, {Finkbeiner}, \&  {Davis} 1998]{sfd98}
{Schlegel}, D.~J., {Finkbeiner}, D.~P., \& {Davis}, M. 1998, \apj, 500, 525

\bibitem[{Schneider}, {Fan}, {Hall}, {Jester},  {Richards}, {Stoughton}, {Strauss}, {SubbaRao}, \& {Vanden Berk} 2003]{dr1q}
{Schneider}, D.~P., {Fan}, X., {Hall}, P.~B., {Jester}, S., {Richards}, G.~T.,  {Stoughton}, C., {Strauss}, M.~A., {SubbaRao}, M., {et al.}, 2003, \aj, 126, 2579 

\bibitem[{Shapley}, {Steidel}, {Adelberger}, \&  {Pettini} 2002]{sha02}
{Shapley}, A.~E., {Steidel}, C.~C., {Adelberger}, K.~L., \& {Pettini}, M. 2002,  in to appear in The New Era of Cosmology, eds. T. Shanks \& N. Metcalfe (San  Francisco: ASP)

\bibitem[{Smith}, {Tucker}, {Kent}, {Richmond},  {Fukugita}, {Ichikawa}, {Ichikawa}, {Jorgensen}, {Uomoto}, {Gunn}, {Hamabe},  {Watanabe}, {Tolea}, {Henden}, {Annis}, {Pier}, {McKay}, {Brinkmann}, {Chen},  {Holtzman}, {Shimasaku}, \& {York} 2002]{sdss105}
{Smith}, J.~A., {Tucker}, D.~L., {Kent}, S., {Richmond}, M.~W., {Fukugita}, M.,  {Ichikawa}, T., {Ichikawa}, S., {Jorgensen}, A.~M., {et al.}, 2002, \aj, 123, 2121

\bibitem[{Steidel}, {Hunt}, {Shapley}, {Adelberger},  {Pettini}, {Dickinson}, \& {Giavalisco} 2002]{sea02}
{Steidel}, C.~C., {Hunt}, M.~P., {Shapley}, A.~E., {Adelberger}, K.~L.,  {Pettini}, M., {Dickinson}, M., \& {Giavalisco}, M. 2002, \apj, 576, 653

\bibitem[{Stoughton}, {Lupton}, {Bernardi},  {Blanton}, {Burles}, {Castander}, {Connolly}, {Eisenstein}, \&  {Frieman} 2002]{sdss85}
{Stoughton}, C., {Lupton}, R.~H., {Bernardi}, M.~B., {Blanton}, M.~R.,  {Burles}, S., {Castander}, F.~J., {Connolly}, A.~J., {Eisenstein}, D.~J., {et al.}, 2002, \aj, 123, 485

\bibitem[{Strauss}, {Weinberg}, {Lupton}, {Narayanan},  {Annis}, {Bernardi}, {Blanton}, {Burles}, {Connolly}, {Dalcanton}, {Doi},  {Eisenstein}, {Frieman}, {Fukugita}, {Gunn}, {Ivezi{\' c}}, {Kent}, {Kim},  {Knapp}, {Kron}, {Munn}, {Newberg}, {Nichol}, {Okamura}, {Quinn}, {Richmond},  {Schlegel}, {Shimasaku}, {SubbaRao}, {Szalay}, {Vanden Berk}, {Vogeley},  {Yanny}, {Yasuda}, {York}, \& {Zehavi} 2002]{str02}
{Strauss}, M.~A., {Weinberg}, D.~H., {Lupton}, R.~H., {Narayanan}, V.~K.,  {Annis}, J., {Bernardi}, M., {Blanton}, M., {Burles}, S., {et al.}, 2002, \aj, 124, 1810

\bibitem[{Tran} 1995]{tran95}
{Tran}, H.~D. 1995, \apj, 440, 597

\bibitem[{van Ojik}, {Roettgering}, {Miley}, \&  {Hunstead} 1997]{vea97}
{van Ojik}, R., {Roettgering}, H.~J.~A., {Miley}, G.~K., \& {Hunstead}, R.~W.  1997, \aap, 317, 358

\bibitem[{Vanden Berk}, {Richards}, {Bauer},  {Strauss}, {Schneider}, {Heckman}, {York}, {Hall}, {Fan}, {Knapp},  {Anderson}, {Annis}, {Bahcall}, {Bernardi}, {Briggs}, {Brinkmann}, {Brunner},  {Burles}, {Carey}, {Castander}, {Connolly}, {Crocker}, {Csabai}, {Doi},  {Finkbeiner}, {Friedman}, {Frieman}, {Fukugita}, {Gunn}, {Hennessy},  {Ivezi{\' c}}, {Kent}, {Kunszt}, {Lamb}, {Leger}, {Long}, {Loveday},  {Lupton}, {Meiksin}, {Merelli}, {Munn}, {Newberg}, {Newcomb}, {Nichol},  {Owen}, {Pier}, {Pope}, {Rockosi}, {Schlegel}, {Siegmund}, {Smee}, {Snir},  {Stoughton}, {Stubbs}, {SubbaRao}, {Szalay}, {Szokoly}, {Tremonti}, {Uomoto},  {Waddell}, {Yanny}, \& {Zheng} 2001]{sdss73}
{Vanden Berk}, D.~E., {Richards}, G.~T., {Bauer}, A., {Strauss}, M.~A.,  {Schneider}, D.~P., {Heckman}, T.~M., {York}, D.~G., {Hall}, P.~B., {et al.}, 2001, \aj, 122, 549

\bibitem[{Vanden Berk et al.} 2004]{vdb04}
{Vanden Berk et al.}, D.~E. 2004, \apj, 601, 692 

\bibitem[{Vernet}, {Fosbury}, {Villar-Mart{\'{\i}}n},  {Cohen}, {Cimatti}, {di Serego Alighieri}, \& {Goodrich} 2001]{ver01}
{Vernet}, J., {Fosbury}, R.~A.~E., {Villar-Mart{\'{\i}}n}, M., {Cohen}, M.~H.,  {Cimatti}, A., {di Serego Alighieri}, S., \& {Goodrich}, R.~W. 2001, \aap,  366, 7

\bibitem[{Villar-Mart{\'{\i}}n},  {Vernet}, {di Serego Alighieri}, {Fosbury}, {Humphrey}, \&  {Pentericci} 2003]{vm03}
{Villar-Mart{\'{\i}}n}, M., {Vernet}, J., {di Serego Alighieri}, S., {Fosbury},  R., {Humphrey}, A., \& {Pentericci}, L. 2003, \mnras, 346, 273

\bibitem[{Visvanathan} \& {Wills} 1998]{vw98}
{Visvanathan}, N. \& {Wills}, B.~J. 1998, \aj, 116, 2119

\bibitem[{Willott}, {Rawlings}, {Jarvis}, \&  {Blundell} 2003]{wrjb03}
{Willott}, C.~J., {Rawlings}, S., {Jarvis}, M.~J., \& {Blundell}, K.~M. 2003,  \mnras, 339, 173

\bibitem[{Yee}, {Ellingson}, {Bechtold}, {Carlberg}, \&  {Cuillandre} 1996]{yea96}
{Yee}, H. K.~C., {Ellingson}, E., {Bechtold}, J., {Carlberg}, R.~G., \&  {Cuillandre}, J.-C. 1996, \aj, 111, 1783

\bibitem[{York}, {Adelman}, {Anderson}, {Anderson},  {Annis}, {Bahcall}, {Bakken}, {Barkhouser}, {Bastian}, {Berman}, {Boroski},  {Bracker}, {Briegel}, {Briggs}, {Brinkmann}, {Brunner}, {Burles}, {Carey},  {Carr}, {Castander}, {Chen}, {Colestock}, {Connolly}, {Crocker}, {Csabai},  {Czarapata}, {Davis}, {Doi}, {Dombeck}, {Eisenstein}, {Ellman}, {Elms},  {Evans}, {Fan}, {Federwitz}, {Fiscelli}, {Friedman}, {Frieman}, {Fukugita},  {Gillespie}, {Gunn}, {Gurbani}, {de Haas}, {Haldeman}, {Harris}, {Hayes},  {Heckman}, {Hennessy}, {Hindsley}, {Holm}, {Holmgren}, {Huang}, {Hull},  {Husby}, {Ichikawa}, {Ichikawa}, {Ivezi{\'c}}, {Kent}, {Kim}, {Kinney},  {Klaene}, {Kleinman}, {Kleinman}, {Knapp}, {Korienek}, {Kron}, {Kunszt},  {Lamb}, {Lee}, {Leger}, {Limmongkol}, {Lindenmeyer}, {Long}, {Loomis},  {Loveday}, {Lucinio}, {Lupton}, {MacKinnon}, {Mannery}, {Mantsch}, {Margon},  {McGehee}, {McKay}, {Meiksin}, {Merelli}, {Monet}, {Munn}, {Narayanan},  {Nash}, {Neilsen}, {Neswold}, {Newberg}, {Nichol}, {Nicinski}, {Nonino},  {Okada}, {Okamura}, {Ostriker}, {Owen}, {Pauls}, {Peoples}, {Peterson},  {Petravick}, {Pier}, {Pope}, {Pordes}, {Prosapio}, {Rechenmacher}, {Quinn},  {Richards}, {Richmond}, {Rivetta}, {Rockosi}, {Ruthmansdorfer}, {Sandford},  {Schlegel}, {Schneider}, {Sekiguchi}, {Sergey}, {Shimasaku}, {Siegmund},  {Smee}, {Smith}, {Snedden}, {Stone}, {Stoughton}, {Strauss}, {Stubbs},  {SubbaRao}, {Szalay}, {Szapudi}, {Szokoly}, {Thakar}, {Tremonti}, {Tucker},  {Uomoto}, {Vanden Berk}, {Vogeley}, {Waddell}, {Wang}, {Watanabe},  {Weinberg}, {Yanny}, \& {Yasuda} 2000]{yor00}
{York}, D.~G., {Adelman}, J., {Anderson}, J.~E., {Anderson}, S.~F., {Annis},  J., {Bahcall}, N.~A., {Bakken}, J.~A., {Barkhouser}, R., {et al.}, 2000, \aj, 120, 1579

\end{thebibliography}


\clearpage

\begin{deluxetable}{cccccccccccc}
\setlength{\tabcolsep}{0.015in}
\tablecaption{Observations of SDSS J113658.36+024220.1\label{t_info}}
\tabletypesize{\scriptsize}
\tablewidth{485.00000pt}
\tablenum{1}
\pagestyle{empty}
\tablehead{
\colhead{MJD of} & 
\colhead{} & \colhead{} & \colhead{} & \colhead{} & \colhead{} & 
\colhead{} & \colhead{} & \colhead{20-cm,} &
\colhead{} & \colhead{Spectroscopic}\\[.2ex] 
\colhead{Observation} & 
\colhead{$u\pm\sigma_u$} & \colhead{$g\pm\sigma_g$} & \colhead{$r\pm\sigma_r$} &
\colhead{$i\pm\sigma_i$} & \colhead{$z\pm\sigma_z$} & 
\colhead{$J\pm\sigma_J$} & \colhead{$K_s\pm\sigma_{K_s}$} & \colhead{mJy} &
\colhead{$M_{i}$} & \colhead{plate/MJD-fiber} 
}
\startdata
51009 & \nodata       & \nodata       & \nodata       & \nodata       & \nodata       & \nodata       & \nodata       & 1.44$\pm$0.15 & \nodata  & \nodata      \\ 
51668 & 22.44$\pm$0.37& 20.69$\pm$0.04& 20.06$\pm$0.03& 19.61$\pm$0.03& 19.28$\pm$0.08& \nodata       & \nodata       & \nodata       & $-$26.30 & \nodata      \\ 
51901 & 21.92$\pm$0.17& 20.54$\pm$0.03& 19.97$\pm$0.02& 19.60$\pm$0.03& 19.12$\pm$0.07& \nodata       & \nodata       & \nodata       & $-$26.31 & \nodata      \\ 
51989 & \nodata       & 20.50$\pm$0.17& 19.86$\pm$0.13& 19.45$\pm$0.11& \nodata       & \nodata       & \nodata       & \nodata       & $-$26.46 & 513/51989-503\\ 
52670 & \nodata       & \nodata       & \nodata       & \nodata       & \nodata       & 18.93$\pm$0.11& 17.77$\pm$0.14& \nodata       & \nodata  & \nodata      \\ 
\enddata
\tablecomments{
Optical magnitudes are SDSS PSF magnitudes \markcite{sdss85}({Stoughton} {et~al.} 2002) except for MJD 51989,
when they are equivalent magnitudes synthesized from the spectrophotometry as
described in \markcite{vdb04}{Vanden Berk et al.} (2004).  To match the optical,
the near-IR magnitudes were measured in 7\farcs4 apertures.
All magnitudes are on the AB system but have not been corrected for Galactic
extinction; the Galactic \ebv\ is 0.02 \markcite{sfd98}({Schlegel}, {Finkbeiner}, \&  {Davis} 1998). 
The 20\,cm column gives the integrated 20\,cm flux density in milli-Janskies
(mJy) from the FIRST survey \markcite{bwh95}({Becker} {et~al.} 1995). 
The absolute magnitude $M_{i}$ was computed assuming $\alpha_{\lambda}=-1.5$
\markcite{dr1q}({Schneider} {et~al.} 2003).  
}
\end{deluxetable}

\begin{deluxetable}{llccrr}
\setlength{\tabcolsep}{0.25in}
\tablecaption{SDSS J1136+0242 Line Detections and Limits\label{t_spec}}
\tabletypesize{\scriptsize}
\tablewidth{440.00000pt}
\tablenum{2}
\pagestyle{empty}
\tablehead{
\colhead{Line}& \colhead{$\lambda_{rest}$\tablenotemark{a}}& \colhead{Redshift}&
\colhead{FWHM\tablenotemark{b}}& \colhead{REW\tablenotemark{c}}&
\colhead{Flux\tablenotemark{d}}
}
\startdata
\multicolumn{6}{c}{Emission Lines}\\
\tableline 
\lya& 1215.6701& 2.4908$\pm$0.0006& 1430$\pm$60& $-$33.8$\pm$3.4& 315$\pm$32\\ 
\Nv& 1240.14& 2.584$\pm$0.002& (1720)& $<-$0.57& $<$5.5\\ 
\oi& 1304.35& 2.483$\pm$0.003& (1430)& $-$1.71$\pm$0.41& 17.4$\pm$4.2\\ 
\cii& 1335.30&  2.500$\pm$0.003& (1430)& $-$1.41$\pm$0.34& 14.7$\pm$3.6\\ 
\SIiv& 1393.755& 2.507$\pm$0.004& (1430)& $<-$1.27& $<$13.8\\ 
\SIiv& 1402.770& 2.507$\pm$0.026& (1430)& $-$0.88$\pm$0.82& 9.6$\pm$8.9\\ 
\civ& 1549.06& 2.496$\pm$0.003& (1510)& $-$1.47$\pm$0.89& 16.7$\pm$10.1\\ 
\heii& 1640.42& 2.495$\pm$0.006& (1430)& $-$0.66$\pm$0.31& 7.5$\pm$3.5\\ 
\Oiii]& 1663.48& 2.478$\pm$0.004& (1430)& $<-$0.32& $<$3.6\\ 
\aliii& 1854.716& 2.499$\pm$0.007& (1430)& $-$0.19$\pm$0.29& 2.1$\pm$3.5\\ 
\aliii& 1862.79& 2.504$\pm$0.002& (1430)& $-$0.70$\pm$0.22& 7.7$\pm$2.4\\ 
\Siiii& 1892.03& 2.500$\pm$0.003& (1430)& $-$0.25$\pm$0.26& 2.7$\pm$2.8\\ 
\ciii& 1908.73& 2.490$\pm$0.003& (1430)& $-$0.33$\pm$0.10& 3.6$\pm$1.1\\ 
\cii]& 2326.44& 2.493$\pm$0.012& (1430)& $-$0.42$\pm$0.36& 4.0$\pm$3.4\\ 
\tableline 
\multicolumn{6}{c}{Associated Absorption Line Systems}\\
\tableline 
\lya& 1215.6701& 2.4589$\pm$0.0011&  510$\pm$160& 1.43$\pm$0.56& \nodata\\
\civ& 1548.202 & 2.4602$\pm$0.0003&  226$\pm$50&  0.84$\pm$0.31& \nodata\\
\civ& 1550.774 & 2.4604$\pm$0.0001&  155$\pm$77&  0.61$\pm$1.29& \nodata\\
\tableline 
\lya& 1215.6701& 2.4771$\pm$0.0012&  910$\pm$150& 7.07$\pm$1.53& \nodata\\
\cii& 1334.5323& 2.4784$\pm$0.0006&  180$\pm$130& 0.63$\pm$0.49& \nodata\\
\SIiv\tablenotemark{e}& 1393.755& 2.4769$\pm$0.0001&  112$\pm$30&  0.27$\pm$0.35& \nodata\\
\SIiv\tablenotemark{e}& 1402.770& 2.4760$\pm$0.0048&  250$\pm$260& 0.55$\pm$0.90& \nodata\\
\civ& 1548.202 & 2.4768$\pm$0.0002&  188$\pm$35&  1.03$\pm$0.22& \nodata\\
\civ& 1550.774 & 2.4769$\pm$0.0003&  212$\pm$71&  0.91$\pm$0.36& \nodata\\
\tableline 
\lya& 1215.6701& 2.4914$\pm$0.0001&  431$\pm$37& 10.1$\pm$1.1& \nodata\\
\SIiv\tablenotemark{e}& 1393.755& 2.4951$\pm$0.0002&   62$\pm$21& 0.18$\pm$0.12& \nodata\\
\SIiv\tablenotemark{e}& 1402.770& 2.4918$\pm$0.0006&  124$\pm$60& 0.26$\pm$0.33& \nodata\\
\civ& 1548.202 & 2.4914$\pm$0.0004&        (431)& 1.12$\pm$0.44& \nodata\\
\civ& 1550.774 & 2.4921$\pm$0.0004&        (431)& 1.19$\pm$0.36& \nodata\\
\tableline 
\multicolumn{6}{c}{Intervening Absorption Line Systems}\\
\tableline 
\civ& 	1548.202& 1.7644$\pm$0.0006&  77$\pm$56& 0.58$\pm$1.10& \nodata\\
\civ& 	1550.774& 1.7637$\pm$0.0003&  97$\pm$37& 0.64$\pm$0.37& \nodata\\
\tableline 
\civ& 	1548.202& 1.8493$\pm$0.0003&  (140$\pm$60)& 1.34$\pm$0.58& \nodata\\
\civ& 	1550.774& 1.8494$\pm$0.0002&  (140$\pm$60)& 1.19$\pm$0.53& \nodata\\
\mgii&	2796.352& 1.8497$\pm$0.0003&   121$\pm$45& 0.83$\pm$0.47& \nodata\\
\mgii&	2803.531& 1.8499$\pm$0.0002&    96$\pm$33& 1.37$\pm$1.29& \nodata\\
\enddata
\tablenotetext{a}{Vacuum rest wavelength in \AA\ \markcite{myj88}({Morton}, {York}, \& {Jenkins} 1988).
For emission lines, we used the $\lambda_{\rm lab}$ wavelengths given in
Table 2 of \markcite{sdss73}{Vanden Berk} {et~al.} (2001), except for cases of well-separated doublets.}
\tablenotetext{b}{Rest-frame FWHM in km s$^{-1}$.  For emission and associated
absorption lines, the rest frame used is that of the quasar ($z=2.4917$).
For intervening absorption line systems, the rest frame is the weighted average
\civ\ redshift of that system.  Parentheses indicate fixed values that were not
allowed to vary during at least the final iteration of the spectral fitting.}
\tablenotetext{c}{Rest-frame equivalent width, in \AA, of the Gaussian fit
to the line.  The rest frames are
the same as those used for the FWHM calculation.  Upper limits are $1\sigma$.}
\tablenotetext{d}{Flux of the Gaussian fit to each emission line,
in units of 10$^{-17}$ ergs cm$^{-2}$ s$^{-1}$.  Upper limits are $1\sigma$.}
\tablenotetext{e}{Only \SIiv\,$\lambda$1402.770 is believably detected at
$z=2.4917$, which is unphysical since it is the weaker of the two lines of
the doublet.  Similarly, \SIiv\,$\lambda$1402.770 appears stronger than 
\SIiv\,$\lambda$1393.755 at $z=2.4768$, which is again unphysical.
Nevertheless, we include these absorption lines in our fit to the spectrum
because doing so yields a more conservative upper limit on the \SIiv\ emission.
No other plausible identifications for these lines have been found (in
particular, we can rule out \NIii\ at both known intervening \civ\ redshifts).}
\end{deluxetable}

\begin{deluxetable}{rlllllllllll}
\setlength{\tabcolsep}{0.04in}
\tablecaption{SDSS J1136+0242 Observed and Model Logarithmic Line Flux Ratios\label{t_ratio}}
\tabletypesize{\scriptsize}
\tablewidth{475.00000pt}
\tablenum{3}
\pagestyle{empty}
\tablehead{
\colhead{} & \colhead{\civ/\lya} & \colhead{\ciii/\lya} & 
\colhead{\Siiii/\lya} & \colhead{\aliii/\lya} & \colhead{\Nv/\lya} &
\colhead{\Nv/\heii}   & \colhead{\Nv/\civ}    & \colhead{\civ/\heii}  &
\colhead{\civ/\ciii}  & \colhead{\ciii/\heii} & \colhead{\cii]/\ciii}
}
\startdata
Data, Variable-$\lambda_{cen}$ & 
$-1.28_{-0.41}^{+0.21}$ & $-1.94_{-0.17}^{+0.12}$ &
$-2.07_{-\infty}^{+0.31}$ & $-1.61_{-0.17}^{+0.12}$ & $<-1.75$ &
$<-0.13$               & $<-0.47$                & $0.35_{-0.63}^{+0.25}$ &
$0.68_{-0.55}^{+0.24}$ & $-0.32_{-0.35}^{+0.19}$ & $0.05_{-1.01}^{+0.28}$ \\
Data, Fixed-$\lambda_{cen}$ & 
$-1.96_{-0.14}^{+0.11}$ & $-2.02_{-0.41}^{+0.21}$ &
$-2.4_{-\infty}^{+0.5}$   & $<-2.03$                & $<-1.7$  &
$<0.21$                & $<-0.74$                  & $-0.05_{-1.38}^{+0.29}$ &
$0.06_{-0.51}^{+0.23}$ & $-0.11_{-\infty}^{+0.21}$ & $0.14_{-0.90}^{+0.27}$ \\
Low Metallicity & 
$-$0.8  & $-$2.35 & $-$2.725 & $-$4.085 & $\simeq-$3.05 &
\nodata & \nodata & \nodata & \nodata & \nodata & \nodata \\
Low Ionization & 
$-$1.65 & $-$1.3  & $-$1.3   & $-$2.4   & $<-$4         &
\nodata & \nodata & \nodata & \nodata & \nodata & \nodata \\
\enddata
\tablecomments{All ``Data'' values are the logarithms of the ratios of the
line fluxes or limits.  Low metallicity ($\log Z/Z_{\odot}=-2$) line ratios
are from Figure 4 of F99
except for \Nv/\lya, which also requires extrapolation from Figure 5 of
\markcite{hea02}{Hamann} {et~al.} (2002). Low ionization ($\log U=-3$) line ratios are from Figure 6 of F99.
Both sets of line ratios assume a power-law continuum with $\alpha_{\nu}=-1.5$
and density $n_H=10^{10}$ cm$^{-3}$; the low-metallicity line ratios also assume
$\log U=-1.5$.
}
\end{deluxetable}

\end{document}